\documentclass[10pt, sigconf]{acmart}

\usepackage{amsmath}
\usepackage{xspace}
\usepackage{algorithm}
\usepackage[noend]{algpseudocode}
\usepackage{tabularx}
\usepackage{array}
\usepackage{calc}
\usepackage{booktabs}
\usepackage{pgfplots}
\usepackage{sansmath}
\usepackage{standalone}
\usepackage{threeparttable}
\usepackage{cleveref}
\usepackage{subcaption}
\usepackage{tikzscale}
\pgfplotsset{compat = 1.14}

\begin{document}
\settopmatter{printacmref=false} 
\renewcommand\footnotetextcopyrightpermission[1]{} 
\makeatletter
\renewcommand\@formatdoi[1]{\ignorespaces}
\makeatother

	\usetikzlibrary{shapes}
	\usetikzlibrary{positioning}
	\usetikzlibrary{matrix}
	\usetikzlibrary{calc}
	\usetikzlibrary{decorations.pathreplacing}
	 \usetikzlibrary{patterns}
	 
	
	\pgfplotsset{compat=1.12}

\pgfplotscreateplotcyclelist{customcyclelist}{%
	densely dashed, blue,  mark=pentagon*\\
	dashed, green, mark=square*\\
	semithick, solid, black, mark=|\\
	semithick, densely dotted, orange, mark=* \\
	semithick, densely dashdotted, red, mark=triangle* \\
	semithick, dashdotdotted, violet, mark=diamond* \\
}

\pgfplotscreateplotcyclelist{customcyclelist_hist}{%
	pattern = north east lines, pattern color=blue\\
	pattern = north west lines, pattern color=green\\
	fill=black\\
	pattern = horizontal lines, pattern color=orange\\
	pattern = crosshatch, pattern color=red\\
	pattern = crosshatch dots, pattern color=violet\\
}
  
  \pgfplotsset{perf plot/.style={
  	cycle list name={customcyclelist},
  	legend style={font=\small},
    label style={font=\small}, 	
    ylabel = Mop/s,
    ymin = 20,
    ymax = 60,
    xmin = 0,
    xmax = 1,
    width=0.35*\linewidth,
    height=0.22*\linewidth
  }}

  \pgfplotsset{single plot/.style={
      width=0.9*\linewidth,
      height=0.50*\linewidth, 
      legend style={font=\small},
      label style={font=\small}, 	
  }}

\pgfplotsset{perfsingle plot/.style={
	perf plot,
    single plot,
	ylabel = Mop/s,
	ymin = 0,
	ymax = 45,
}}

  \pgfplotsset{perf load plot/.style={
    perf plot,
    xmin = 0.6,
    xmax = 0.95,
    xlabel=load factor,
  }}

  \pgfplotsset{perf negative plot/.style={
    perf plot,
    xmin = 0.0,
    xmax = 1.0,
    xlabel=negative lookup rate,
  }}

\pgfplotsset{perf plot insert/.style={
		perf load plot,
		ymin= 0,
		ymax= 25,
		xtick=data,
		ybar = 0pt,
		bar width = 4pt,
		xmin=0.5,
		xmax=1.1,
		cycle list name={customcyclelist_hist},	
}}

\settopmatter{printfolios=true}
\setcopyright{none}%
\title[Cuckoo++ Hash Tables]{Cuckoo++ Hash Tables: High-Performance Hash Tables for Networking Applications}

\acmConference[Technicolor Technical Report]{Technicolor Technical Report}{December 2017}{Rennes, France}                                                 
\acmYear{2017} 

\author{Nicolas 
	Le Scouarnec}

\affiliation{%
	\institution{Technicolor}                                                                                  
}                                                                                                                                       

\sloppy

\begin{abstract}
	Hash tables are an essential data-structure for numerous networking applications (e.g., connection tracking, firewalls, network address translators). Among these, cuckoo hash tables provide excellent performance by allowing lookups to be processed with very few memory accesses (2 to 3 per lookup). Yet, for large tables, cuckoo hash tables remain memory bound and each memory access impacts performance. In this paper, we propose algorithmic improvements to cuckoo hash tables allowing to eliminate some unnecessary memory accesses; these changes are conducted without altering the properties of the original cuckoo hash table so that all existing theoretical analysis remain applicable.	 
	On a single core, our hash table achieves 37M lookups per second for positive lookups (i.e., when the key looked up is present in the table), and 60M lookups per second for negative lookups, a 50 \% improvement over the implementation included into the DPDK. On a 18-core,  with mostly positive lookups, our implementation achieves 496M lookups per second, a 45\% improvement over DPDK.
\end{abstract}

\maketitle

\section{Introduction}
The increasing I/O performance of general purpose processors (a dual-socket Xeon can accommodate up to 10 40Gbps network interface cards) as well as the availability of frameworks for high-performance networking (DPDK~\cite{DPDK}, Netmap~\cite{Rizzo2012}, PFQ~\cite{Bonelli2016}...), allow replacing hardware-based network middleboxes support by commodity servers. This trend is supported by systems relying on software-based implementation of network functions~\cite{Palkar2015, Panda2016, Zhou2013} that target performance in the order of millions packets per seconds (Mpps) per core. The applications running on top of these include L2 (switches), L3 (routers) and L4 (load balancing, stateful firewalls, NAT, QoS, traffic analysis). In order to support L4 applications, an important feature is the ability to identify connections/flows and to keep track of them.

More specifically, these applications require to associate some state to each connection. For a load-balancer, the state is the destination server to use; for a firewall the state specifies whether the connection is allowed or not; for a NAT, the state is the addresses and ports to use when  translating from one network to another; and for QoS or traffic analysis the state can contain packet counters. In the case of IP protocols (UDP/TCP over IP), the connection tracking is achieved by identifying the connection using its 5-tuple (protocol, source address, destination address, source port and destination port) and mapping an application-specific value to this 5-tuple. As each connection has its own state, the number of entries in the system grows with the number of flows. As each route sees many concurrent flows going over it, the scale is much higher: when routing and forwarding require to consider tens of thousands of routes (L3 information), connection tracking requires to track millions of flows. To deal with these requirements, the association is often stored in a hash table to access it efficiently. 

High-performance hash tables often rely on bucketized cuckoo hash-table~\cite{Pagh2004, Dietzfelbinger2007, Breslow2016, Fan2013, Li2014, Panigrahy2005, Zhou2013} for they feature excellent read performance by guaranteeing that the state associated to some connection can be found in less than three memory accesses. Bucketized cuckoo hash tables are open-addressed hash tables where each value may be stored into any slot of two buckets determined by hashing the key. When used for networking applications (i.e., packet processing), an important feature of these hash tables is their ability to perform lookups in batches as supported by CuckooSwitch~\cite{Zhou2013} or DPDK's implementation~\cite{DPDK} as it allows to efficiently prefetch the memory accesed for processor efficiency. As we discuss in Section~\ref{sec:study}, two implementation choices are possible regarding these prefetches: \emph{(i)} optimistically assume that the data will be in the primary bucket and prefetch only this bucket as done in Cuckoo Switch~\cite{Zhou2013}, or (ii) pessimistically assume that both buckets need to be searched and prefetch both to avoid too late prefetching and mis-predicted branches as done in DPDK~\cite{DPDK}. Yet, as we show in Section~\ref{sec:study}, none of these two strategies is optimal for all situations. Moreover, this becomes more problematic as boxes get exposed to uncontrolled traffic (e.g., NAT or firewall exposed on the Internet are likely to receive packets of unknown flows that should be filtered or could be subject to a DoS attempt).

In this paper, we describe algorithmical changes to cuckoo hash tables allowing a more efficient implementation. More precisely, our hash table adds a bloom filter in each bucket that allows to prune unnecessary lookups to the secondary bucket. It has the following features: \emph{(i)} high performance for both positive and negative lookups with up to 37-60 Millions lookups per second per core, so that performance does not decrease with malicious or undesired traffic, and \emph{(ii)} builtin timers to support the expiration of tracked connections.

Section 2 describes the background on packet processing applications and their need in term of data-structure for connection tracking; as well as the state of the art on cuckoo hash tables. Section 3 studies the implementation choices of cuckoo hash tables and analyze the performance of the various alternatives. In Section 4, we introduce Cuckoo++ hash tables, an improvement over bucketized cuckoo hash tables that allows for universally more efficient implementation. In Section 5, we evaluate their performance. In Section 6, we review related work and we conclude in Section 7.

\section{Background}
\label{sec:back}
Packet-processing networking applications are diverse: beside switching and routing well supported by standard networking hardware, they include more advanced applications such as traffic analysis (IDS/IPS), firewalling, NATting, QoS, or L4 load balancers. As a common characteristic, these applications often require to keep connection-related state.  

These applications must store a very large number of entries (i.e., one per connection or flow), thus requiring a large amount of memory (e.g., 32M flow entries of 256 bits require 1GB of memory). ASICs dedicated to high-performance networking fail to meet these requirements as they are much more limited in capacity : the largest TCAMs available today are 80 Mb (or 10 MB) and Network Search Processors that build on SRAM are limited to 1 Gb (or 128 MB). Hence, while these ASICs are well-suited for L2/L3 processing (i.e., MAC learning or IP routing), software running on commodity CPUs (e.g., Intel Xeon) that can address large amount of DRAMs is a cost-efficient alternative for high-performance L4-L7 packet processing requiring connection tracking.

Software-based implementation of high-performance packet processing is supported by the availability of kernel bypass solutions~\cite{DPDK,Bonelli2016,Rizzo2012} supporting millions of packets per second per core. On a dense system, it is possible to put up to 10 40 Gbps cards with 2 Intel Xeon processors (20-40 cores). Given standardly sized packets (i.e., Simple IMIX), it means that 3.6-7.2 millions packets must be processed per second per core. This requires highly optimized implementations, in which \emph{(i)} the kernel is bypassed to avoid overhead, \emph{(ii)} I/O costs are amortized by sending and receiving batch of packets (e.g., in DPDK the application receives batches of 32 packets from the NIC), \emph{(iii)} a share-nothing architecture is adopted to avoid synchronization overheads, and \emph{(iv)} highly efficient algorithms and data-structures are used. 

In share-nothing architecture, the network interface card (NIC) steers the packets to several queues. Each core reads and sends through its own set of queues; it shares no data-structure with other cores to avoid sharing and synchronization (even implicit) which reduces performance. This allows good scaling properties. Share-nothing architecture are well supported by modern fast NIC that provide several facilities for distributing packets to queues in a flow-coherent way (e.g., RSS, Intel Flow Director, ...). As a consequence, data-structure for high performance packet processing applications don't need to support multiple writer/multiple reader. This allows to simplify them and focus on improving their performance in a mono-threaded setting.

\begin{figure}[t]
	\centering
	\subcaptionbox{small keys\label{fig:layout_small}}{%
		\hbox{
				\hfill
		\includegraphics[height=0.30\textwidth]{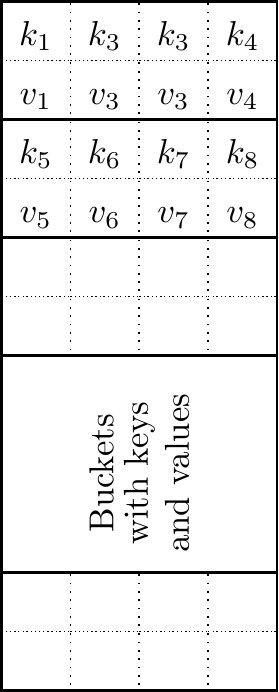}
	\hfill}}%
	\hfill%
	\subcaptionbox{larger keys and values		\label{fig:layout_large}}{%
		\includegraphics[height=0.30\textwidth]{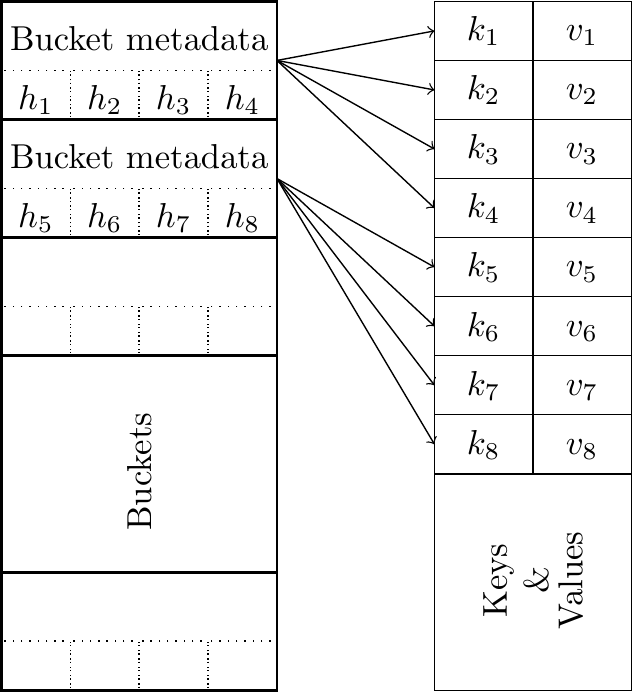}}%
	\caption{Memory layout for cuckoo hash table}
\end{figure}

In software, the standard approach to store per-connection state is to use a hash table. High-performance hash tables often rely on open addressing hash schemes~\cite{Celis1985,Herlihy2008,Pagh2004, Panigrahy2005}. Open addressing hash tables avoid pointer chasing, that generates many costly memory accesses. Among these, cuckoo hash tables~\cite{Pagh2004, Panigrahy2005} allow lookups to be performed in 2 to 3 memory accesses. Cuckoo hash tables achieve these results by moving complexity from lookup to insertion. The insertion cost is thus increased, but this strategy is beneficial as lookups tend to dominate execution time. Consequently, many high performance hash tables~\cite{Breslow2016,Fan2013,Li2014,DPDK,Zhang2015,Zhou2013} now implement bucketized cuckoo hash tables~\cite{Dietzfelbinger2007} with 2 hashes as this variant allows improved performance in practice and higher load factors. These implementations target different context: some were designed for GPUs~\cite{Breslow2016}, others are a shared data-structure for applications such as key-value stores~\cite{Fan2013,Li2014,Zhang2015}. In high-performance networking, a desirable feature is batched lookups as supported by CuckooSwitch~\cite{Zhou2013} and DPDK~\cite{DPDK}. Batched lookups match the execution model of typical DPDK programs and allow optimized implementation with improved performance on out-of-order super-scalar CPUs (e.g., Intel Xeon): the different steps of the individual lookups that compose the batch are interleaved so as to hide memory latency and increase the instruction level parallelism

In bucketized cuckoo hash table, the memory is divided in buckets of fixed size (4 or 8 slots): each bucket is sized to fit a cacheline ( i.e., the unit at which the processor accesses memory). Each key hashes to a primary bucket (i.e., the bucket indexed by the primary hash) and a secondary bucket (i.e., the bucket indexed by the secondary hash). The value associated to a key must be stored in one of the slots of these two buckets. Two implementations choices are possible for storing the values: \emph{(i)} if the key and values are small enough (e.g., less than 64 bit total), they can be stored directly into the bucket as shown on Figure~\ref{fig:layout_small}; \emph{(ii)} larger keys and values do not allow the bucket to remain on a single cacheline, the solution is thus to store only the hashes and an index/pointer in the bucket : this allows the lookup to access the bucket efficiently to find the index, and once the index is found one additional memory access is necessary to obtain the key and value, as shown on Figure~\ref{fig:layout_large}. As our focus is on data structures for connection tracking (128 bit keys), we do not consider the restricted case of small keys and values and thus adopt the second approach. 

Using this data structure, looking up the value associated to a key is immediate. The key is hashed, the two associated buckets are accessed, the hash is compared to the hashes stored into the slots of the buckets, and if some hash matches, the corresponding entry in the key/value array is accessed. Since the matching hash may be a false positive, the key of the entry is compared to the searched key. If they match, the lookup answers positively and the value is returned. If no hash matches, or if no key matches, then the lookup answers negatively. Thus, at most, 3 memory accesses (i.e., to the primary bucket, to the secondary bucket and to the key and value) are needed to answer lookups.

This efficient lookup procedure is enabled by the cuckoo insertion procedure that can re-structure the hash table to ensure than an entry can always be inserted in either its primary or its secondary bucket. If a free slot can be found in the primary or secondary bucket, the hash is written to the slot and the key/value is written at the corresponding index in the key/value array. If no free slot is found (i.e., in case of collision) the cuckoo insertion procedure makes up a free slot by moving one of the entry to its alternative bucket (i.e., to its secondary bucket if it is stored in the primary bucket, or the inverse). This procedure may apply recursively to make up free space for the moved entry. This set of exchanges between primary buckets and secondary buckets form a cuckoo path, and is what allows cuckoo hash table to achieve high load factor while guaranteeing that during lookup only two buckets need to be read.  

Improving lookup performance is a recurring concern in open addressing hash tables~\cite{Celis1985,Herlihy2008,Pagh2004}. The goal is to minimize the number of memory accesses for all cases. Generally, negative lookups are the worst case for it requires to check all possible positions to ensure that it is not stored at any position. On the contrary, positive lookups can stop early without checking all positions as soon as the key is found. Cuckoo hashing have been developed precisely to ensure that even for the worst case the number of memory accesses is small and bounded. While these theoretical developments ensure that the number of memory accesses remains low, they do not exploit the features of modern hardware. For instance, modern processors access memory by chunks of 64 bytes (known as cachelines) and the memory latency varies heavily depending on the ability to predict long in advance memory accesses so as to prefetch the corresponding memory location. Taking these features into account leads to large performance gains. In  Section~\ref{sec:study}, we evaluate the performance of optimized implementations that target modern processors. 

In the rest of this paper, we focus on hash tables targeted at use in packet processing applications having the following characteristics: \emph{(i)} high performance on commodity (i.e., Intel Xeon) processors (10s of millions of lookups per second), \emph{(ii)} support for lookups in batches, as DPDK applications process batches of packets from the hardware, \emph{(iii)} support for connection tracking applications (i.e., 128-bit keys and values).
The only implementation available supporting these requirements is the cuckoo hash table implemented in DPDK. As of version 17.05, DPDK implements a bucketized cuckoo hash table, leveraging SIMD, having 8-slots per bucket, supporting arbitrarily sized keys and 64-bit values. Note that our context slightly differs from CuckooSwitch~\cite{Zhou2013} or~\cite{Li2014} for the use of share-nothing architecture allows to avoid the complexity and overheads related to synchronization

\section{Implementing Cuckoo Tables}
\label{sec:study}
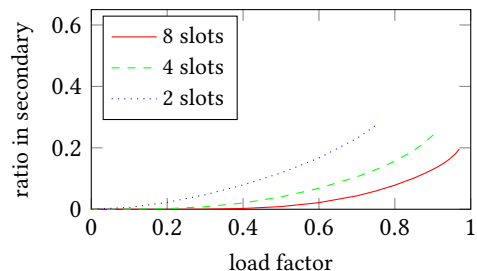
\begin{figure}[bt]
	\centering
	\begin{tikzpicture}
\begin{axis} [
perfsingle plot,
cycle list={black},
legend pos=north west,
ylabel= ratio in secondary,
ymin= 0,
ymax= 0.65,
xmin= 0,
xmax= 1,
xlabel= load factor,
unit vector ratio=1 0.9,
legend columns = 1
]

  \addplot+ [solid, color=red, mark=]
    table[x=filling_ratio, y=stats_in_secondary] {plots/ratio_in_secondary_8_bloom.dat};
  
  \addplot+ [dashed, color=green, mark=]
    table[x=filling_ratio, y=stats_in_secondary] {plots/ratio_in_secondary_4_bloom.dat};
  
  \addplot+ [dotted, color=blue, mark=]
	table[x=filling_ratio, y=stats_in_secondary] {plots/ratio_in_secondary_2_bloom.dat};


     \legend{8 slots\\ 
     	     4 slots\\
       	     2 slots\\}
\end{axis}
\end{tikzpicture}
	\caption{Fraction of entries stored in their secondary bucket for a 32M-capacity cuckoo hash table.}
	\label{plots:ratio_in_secondary}
\end{figure}

All high-performance implementations~\cite{Zhou2013,Fan2013, Li2014, DPDK, Breslow2016} rely on bucketized cuckoo hash table~\cite{Dietzfelbinger2007} that have multiple slots for storing entries (e.g., 4 or 8) in each bucket (Section~\ref{sec:back}). The lookup procedure can stop early when it finds the key: the secondary bucket needs to be accessed only when the key is stored in it, or for negative lookups. Figure~\ref{plots:ratio_in_secondary} plots the fraction of entries stored into their secondary bucket; the curve stop on the x-axis at the load factor at which insertions start to fail. Even for high-load factors, most entries (i.e., more than 80\%) remain stored in their primary bucket when using 8-slots per buckets as in~\cite{DPDK, Li2014}: when the hash table is filled to less than 50\%, almost no entry are in their secondary bucket, at 70\% of their capacity, 6\% of entries are in their secondary bucket and at 95\%  of its capacity, only 16\% of the entries are in their secondary bucket. This is an improvement over designs with 4 slots per bucket~(e.g., CuckooSwitch~\cite{Zhou2013}, MemC3~\cite{Fan2013}, DPDK prior to version 16.10~\cite{DPDK}). Also note that 1 or 2 slots per bucket are impractical for they don't allow high load factors without insertion errors in practical settings.

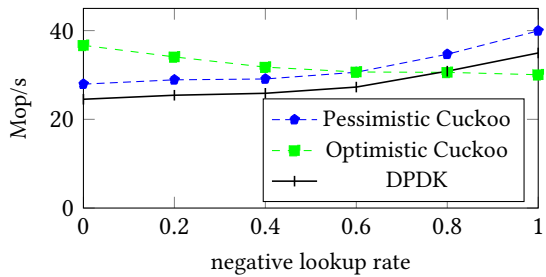
\begin{figure}[t]
	\centering
%
%
%
\begin{tikzpicture}[baseline]
\begin{axis} [
  perfsingle plot,
  legend pos= south east,
  xlabel=negative lookup rate,
]

\addplot+
table[x=invalid_rate, y=UNCOND_op_rate] {graphs/x_invalid-load0.8.gnuplot};
\addplot+
table[x=invalid_rate, y=COND_op_rate] {graphs/x_invalid-load0.8.gnuplot};
\addplot+
table[x=invalid_rate, y=DPDK_1702_op_rate] {graphs/x_invalid-load0.8.gnuplot};

\legend{Pessimistic Cuckoo\\
  Optimistic Cuckoo\\
  DPDK\\}

\end{axis}
\end{tikzpicture}
	\caption{Performance of batched lookup (32M-capacity cuckoo hash table, load factor of 0.8).}
	\label{plots:perf_analysis}
\end{figure}

For super-scalar out-of-order CPUs (e.g., Intel Xeon), the performance of implementations of cuckoo hash tables that support batched lookups highly depends on the effectiveness of prefetches\footnote{Prefetch is an instruction that can be issued to inform the processor that some memory location will be accessed in the future. It allows the processor to "pre-load" a specific memory location into its cache thus avoiding long memory latencies.} that hide the memory latency~\cite{Zhou2013, DPDK}. Other implementations, which do not support batching, are less impacted by the prefetches as there are less opportunities for hiding memory latency on a single lookup. Two prefetching strategies are possible for implementing cuckoo hash tables.

The \emph{optimistic} approach assumes that the key is stored in the hash table and that it will be in the primary bucket: thus only the primary bucket is prefetched, the secondary bucket will be prefetched later only if the data is not found in the primary bucket. This optimistic approach is adopted by CuckooSwitch~\cite{Zhou2013}, MemC3~\cite{Fan2013}, Mega-KV~\cite{Zhang2015}. This approach is supported by the previous observation (see Figure~\ref{plots:ratio_in_secondary}) that less 20\% positive lookups need to access the secondary bucket. This saves resources at the price of a few non-prefetched memory accesses.

The \emph{pessimistic} approach assumes that the key is not in the hash table or that it could be in the secondary bucket: thus both the primary and the secondary bucket are prefetched. Beside hiding memory latency, this approach also avoids branch mispredictions, which are detrimental to performance in out-of-order CPUs. This pessimistic approach is adopted in DPDK~\cite{DPDK} and \cite{Li2014, Ross2007}.

Intuitively, one can observe that this choice is highly dependent on the workload: the optimistic choice should be favored for mostly-positive lookups, and the pessimistic choice should be favored for mostly-negative lookups. 

To study the actual impact on performance of these approaches, we implement optimistic cuckoo hash table and pessimistic cuckoo hashtable with exactly the same memory layout which is detailed in Section~\ref{sec:main}. The two implementations share a common code base which is highly optimized and supports batching, a key enabler for high performance in packet processing applications. Note that we can't directly rely on CuckooSwitch~\cite{Zhou2013}, and DPDK~\cite{DPDK} for they differ greatly in what they support (e.g., size of keys, SIMD, hash function, batching). Indeed, CuckooSwitch is tailored for L2 switching and thus only supports 48-bit keys (i.e., MAC addresses) and 16 bit values (port), DPDK supports larger keys and batching but implements only the pessimistic approach. To showcase the performance of our implementation, we use DPDK's highly-optimized cuckoo hash table as a baseline.

Figure~\ref{plots:perf_analysis} plots the number of lookups per second achieved for batched lookups on a hash table with a capacity of 32M entries using CityHash 64~\cite{Pike2011} hash function. We vary the load factor, and the ratio of negative lookups. As expected, our pessimistic approach (pessimistic cuckoo) exhibits a similar behavior to DPDK (which is also a pessimistic approach). Our implementation slightly exceeds DPDK's performance. 

\begin{figure*}[t]
	\centering
	\centering
	\subcaptionbox{Positive lookup\label{fig:overall_positive}}{
		\includegraphics[width=0.32\linewidth]{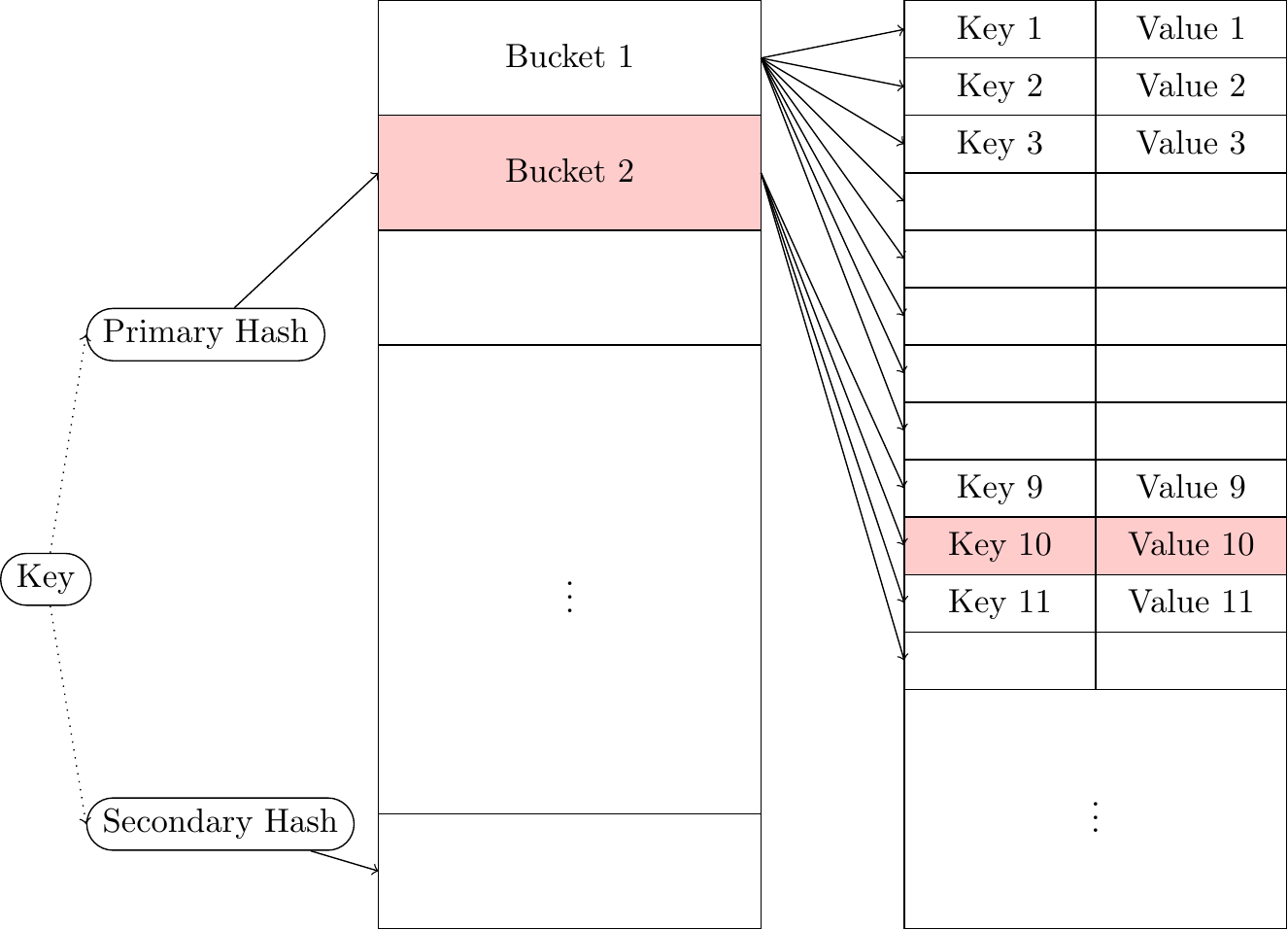}}%
	\hfill%
	\subcaptionbox{Negative lookup (Cuckoo) \label{fig:overall_neg_cuckoo}}{
		\includegraphics[width=0.32\linewidth]{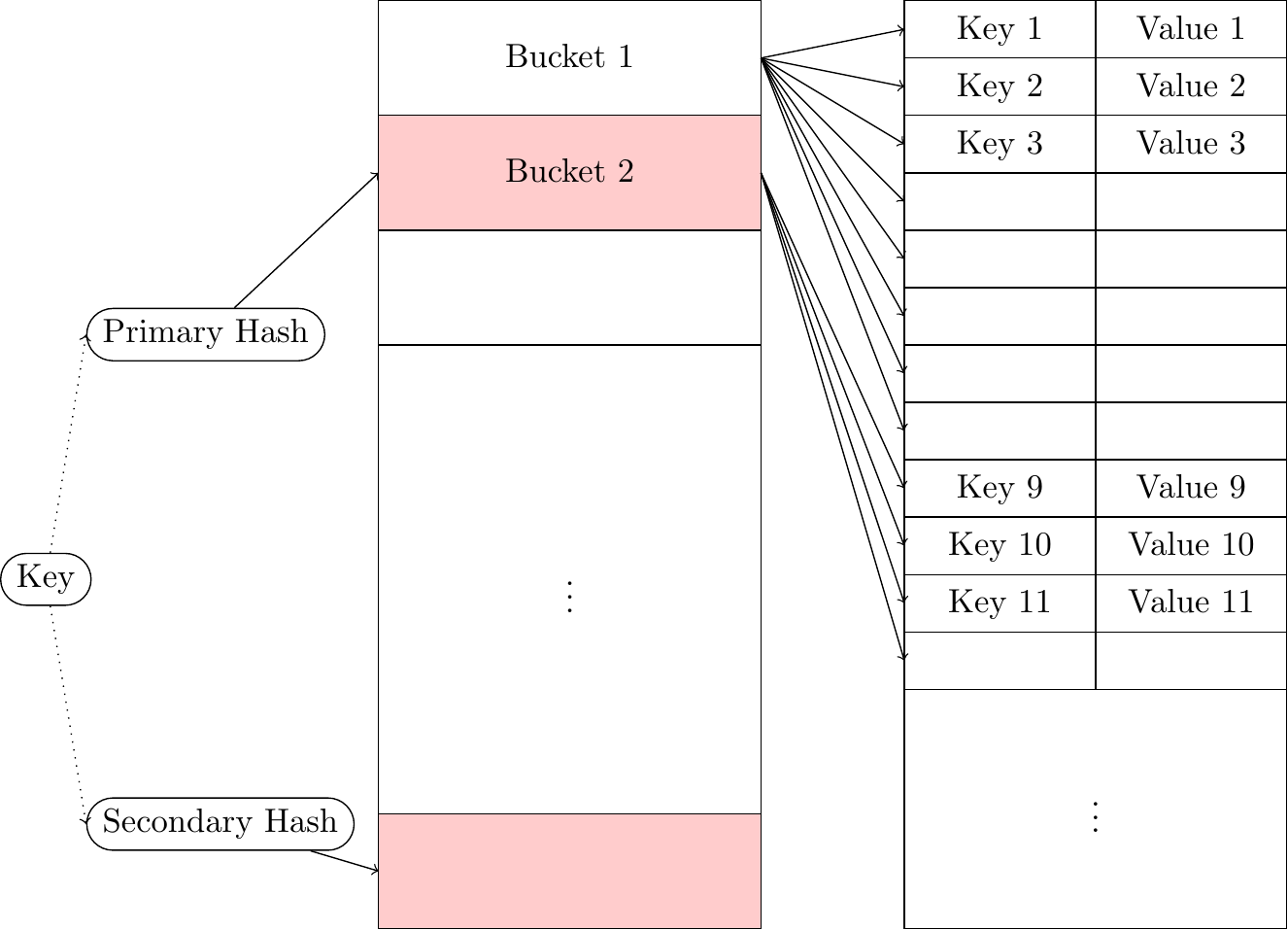}}%
	\hfill%
	\subcaptionbox{Negative lookup (Cuckoo++) \label{fig:overall_neg_cuckooplus}}{
		\includegraphics[width=0.32\linewidth]{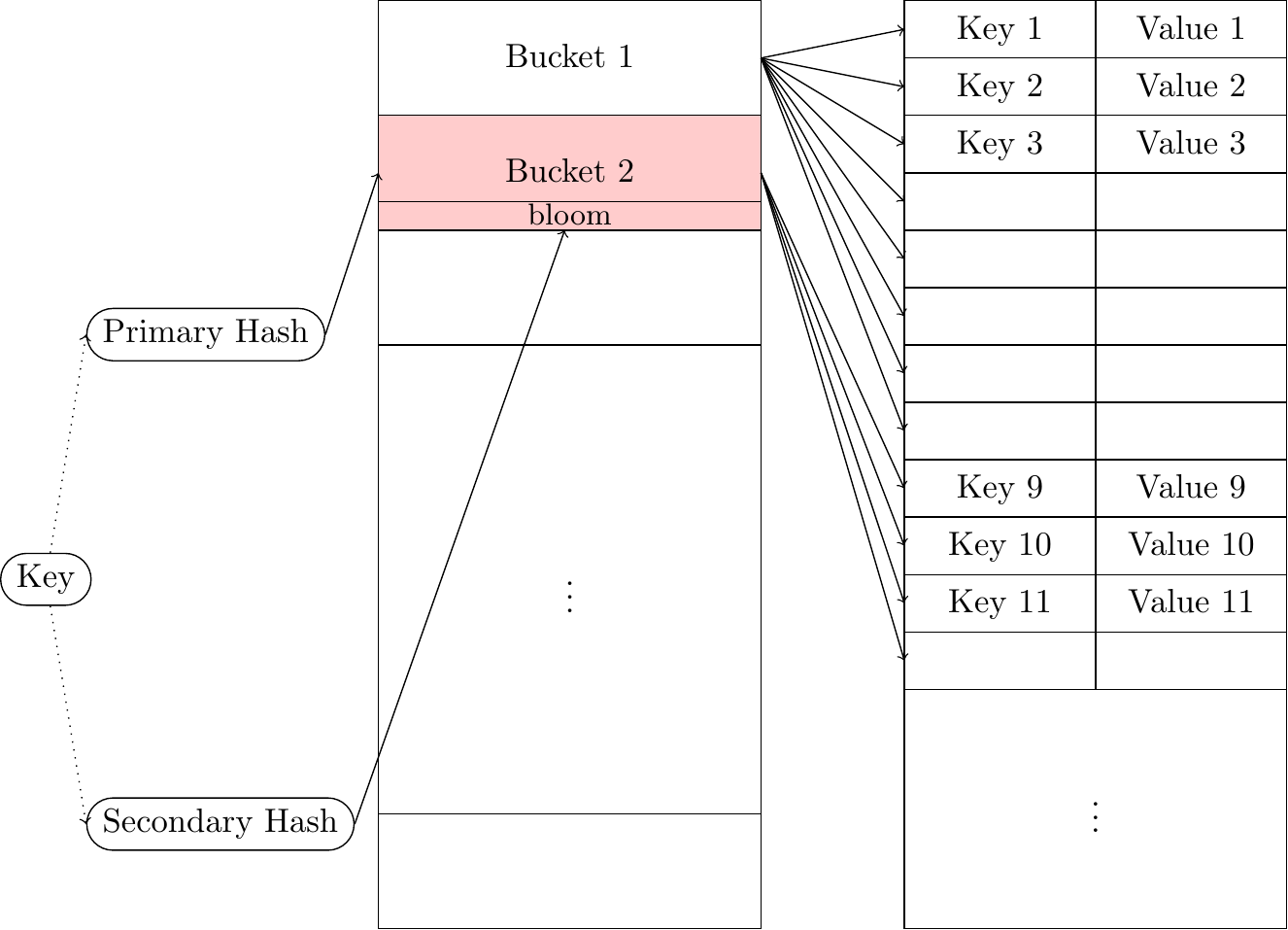}}%
	\caption{Overall organization of the hashtable. The memory accessed during lookups is shaded.}
	\label{fig:overall}
\end{figure*}

Optimistic cuckoo outperforms pessimistic cuckoo when performing mostly positive lookups. This shows that it is unnecessary and detrimental to systematically prefetch the secondary bucket as, even for high load factors, entries are stored into their secondary buckets in less than 20\% of cases. By contrast, when performing mostly negative lookups, pessimistic cuckoo outperforms optimistic cuckoo as prefetches are more efficient for they are issued in advance. 

As a consequence, no implementation can be deemed as ideal since the optimistic implementation will perform better when most lookups are successful while the pessimistic implementation will perform better in the opposite case. Hence, a connection tracking system with an optimistic implementation will have a higher performance, yet its performance will degrade in presence of invalid or malicious traffic.

Thus, packet processing libraries need to offer several implementations of hash tables, and users need to know the properties of each implementation to choose the one that will perform best on their workload. This is not practical as the workload may not be precisely characterized or may be dynamic: for example, a firewall/NAT system could see a spike in unknown traffic or a transiently increased number of connections (e.g., DoS attack). It is crucial that the performance of these network functions does not degrade significantly when facing unusual or malicious traffic.

Our paper addresses this issue. To this end, we propose algorithmic changes to bucketized cuckoo hash tables that enable implementations offering a high performance in all cases (i.e., all load factors and all negative lookup rates). These algorithmic changes have as little impact on baseline performance as possible so they can always be chosen in place of standard bucketized cuckoo hash table.

\section{Cuckoo++ Hash Tables}
\label{sec:main}

Cuckoo++ hash tables build upon standard cuckoo hash tables: through this design choice, all theoretical guarantees provided for cuckoo hash tables apply to Cuckoo++ hash tables. Yet, to meet the requirements of high performance networking, Cuckoo++ offers several improvements that allow avoiding:
\emph{(i)} many memory accesses to the secondary bucket during lookups for improved performance (Section~\ref{ssec:avoid}),
\emph{(ii)} the overhead of triggering timeouts by implementing timers directly into the hash table and the overhead of memory accesses for deletion by relying on lazy deletion (i.e., deleting only when an expired entry is accessed) (Section~\ref{ssec:timer}). 
All these algorithmic changes are done under the constraint of an efficient execution on general purpose processors by allowing an optimized memory layout (Section~\ref{ssec:memory}).

\subsection{Avoid accesses to the secondary bucket}
\label{ssec:avoid}
One of the main design goal of Cuckoo++ is to minimize the number of cachelines accessed as the memory bandwidth is a scarce resource that must be used parsimoniously. The improvement over standard Cuckoo hash tables is illustrated on Figure~\ref{fig:overall}. This Figure depicts a typical hash table with a first array of buckets storing metadata about entries and a second array storing the complete keys and values.

For a positive lookup (Figure~\ref{fig:overall_positive}), for both Cuckoo and Cuckoo++ hash table, the key is hashed and the primary bucket is accessed. The searched key is found in the bucket and the index in the second array is determined. The value is then accessed and recovered, thus answering most lookups with only 2 memory accesses. In some cases, the secondary bucket must be accessed but this remains rare.

For a negative lookup, Cuckoo and Cuckoo++ differ. For standard cuckoo hash tables (Figure~\ref{fig:overall_neg_cuckoo}), the key is hashed, the entry is not found in the primary bucket, the secondary bucket needs to be checked before deciding that they searched key is not stored in the hash table. This second memory access tends to be costly as it cannot be predicted in advance (i.e., only after the searched key is not found in the primary bucket). Thus, negative lookups require 2 memory accesses. For Cuckoo++ (Figure~\ref{fig:overall_neg_cuckooplus}), we add a bloom filter in the metadata of the primary bucket. This bloom filter contains all keys that could not be stored into this primary bucket and that have been moved to their secondary bucket. During lookup, before fetching the secondary bucket from memory, the searched key is searched in the bloom filter: two hashes are derived from the secondary hash of the key/value, and the bits of the bloom filter indexed by these hashes are tested. If the bloom filter answers negatively (\emph{i.e.,} at least one of the bit is zero), then the secondary bucket will not contain the searched key. If the bloom filter answers positively (\emph{i.e.,} all bits are ones), then the secondary bucket is fetched: it may or may not contain the key since bloom filters have false positives. Thus, the bloom filter acts as a hint that allows to determine if a key could have been stored in its secondary bucket. Thus, accesses to the secondary bucket can be avoided in most cases ensuring that most negative lookups can be answered in 1 memory access.

When a key and value is stored into its primary bucket, the insertion procedure is identical to standard cuckoo hash table. When a key and value is stored in its secondary bucket, in addition to the insertion procedure of standard cuckoo hash tables, the bloom filter of the corresponding primary bucket must be updated. Two hashes are derived from the secondary hash of the key/value to be inserted, and the bits of the bloom filter indexed by these hashes are set. A counter is associated to each bloom filter; it counts the number of insertions in the bloom filter and is thus incremented for each insertion in the bloom filter.

  \begin{figure}[b]
 	\centering
 	\includegraphics[width=0.9\linewidth]{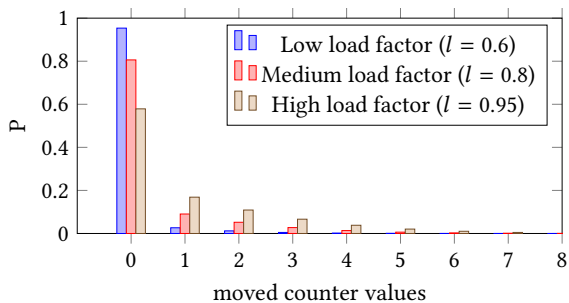}%
 	\caption{Distribution of values of Moved Counter for a 32M-capacity Cuckoo++ hash table.}
 	\label{fig:distrib_moved}
 \end{figure}

When deleting a key from the hash table, the usual cuckoo procedure applies. When the key was stored into its secondary bucket, the bloom filter of the corresponding primary bucket should be updated. Yet, bloom filters are append-only structures and values added to bloom filters cannot be removed. To deal with this constraint, we leverage the counter associated to each bloom filter. The counter is decremented for each deletion of a value stored in its secondary bucket. Whenever the counter reach zero, we can reset the bloom filter to its empty state (i.e., reset all bits to zero).

When collisions must be resolved during an insertion, a chain of swaps (i.e., a cuckoo path) is computed. Key/values moved from their primary bucket to their secondary bucket modify the bloom filter similarly to an insertion, whereas keys/values moved from their secondary bucket to their primary bucket modify the bloom filter similarly to a deletion.

In practice, most entries are stored in their primary position (see Figure~\ref{plots:ratio_in_secondary}), it implies that occupancy of bloom filters remains low and that moved counter values remain low (see Figure \ref{fig:distrib_moved}). As a consequence, over the lifetime of the hash table, the counter is often equal to zero, ensuring that the bloom filter is often reset and remains useful. Note that even for high load factor (0.95) more than 50\% of buckets have a moved counter value equal to zero. 

Regarding the efficiency of the bloom filter for avoiding accesses to the secondary bucket, we use a bloom filter of 64 bits, with two hash functions. For such a bloom filter, given the number of entries inserted in the bloom filter (see Figure~\ref{fig:distrib_moved}), the false positive rate of the bloom filter remains very low even for high load factor (e.g., 0.003 for a load factor of 0.95 -- other load factors are given in Table~\ref{tab:fpr}).

\begin{figure*}[t]
	\centering
	\setbox9=\hbox{\includegraphics[width=0.32\textwidth]{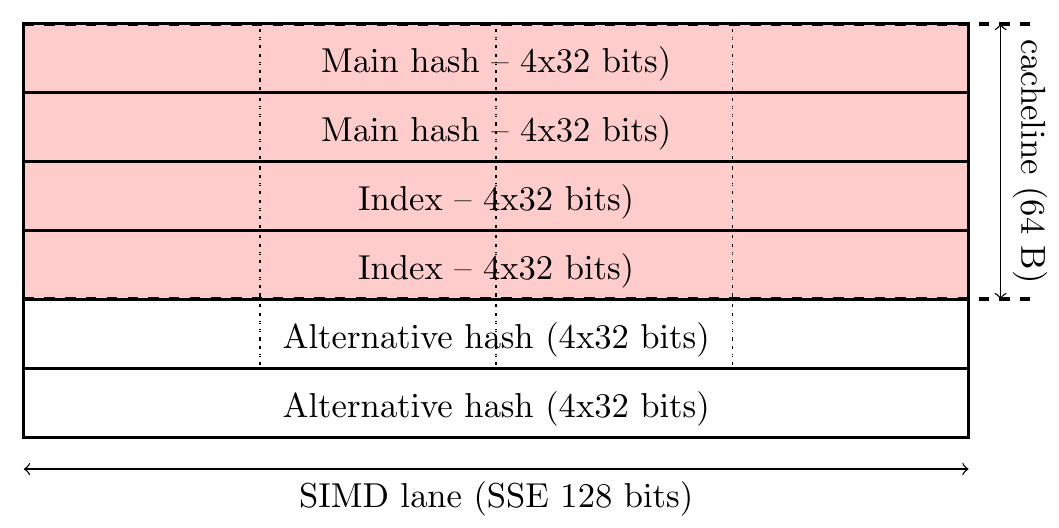}}%
	\subcaptionbox{DPDK's layout}{%
		\raisebox{\dimexpr\ht9-\height}{\includegraphics[width=0.32\textwidth]{fig/memory-dpdk.pdf}}}%
	\hfill%
	\subcaptionbox{Cuckoo++ layout (without timers)}{%
		\raisebox{\dimexpr\ht9-\height}{\includegraphics[width=0.32\textwidth]{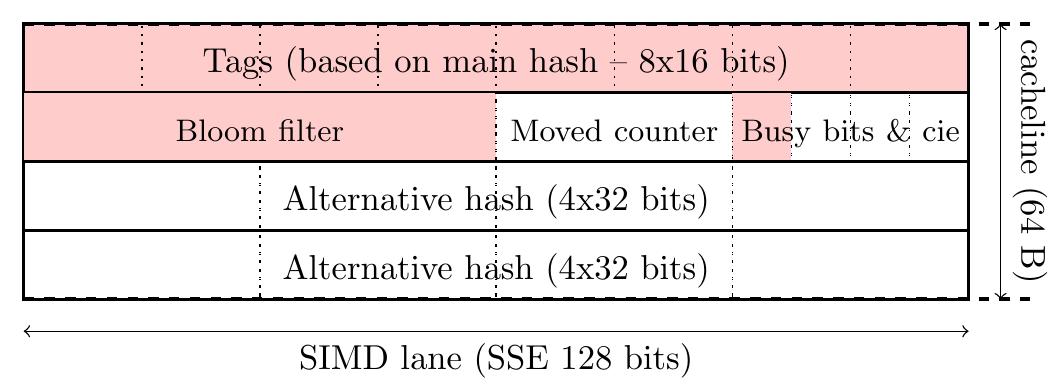}}}%
	\hfill%
	\subcaptionbox{Cuckoo++ layout (with timers)}{%
		\raisebox{\dimexpr\ht9-\height}{%
			\includegraphics[width=0.32\textwidth]{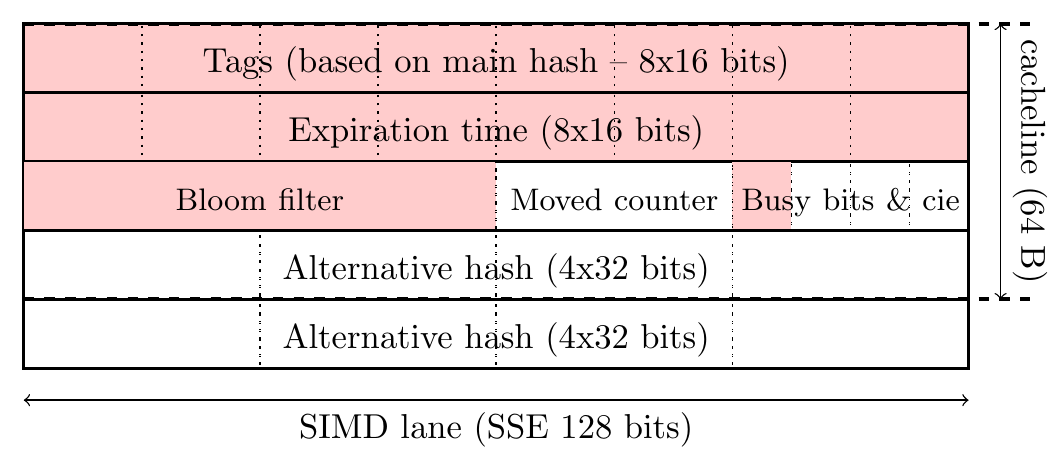}}}%
	\caption{Bucket Memory layout}
	\label{fig:bucket_memory}
\end{figure*}

\subsection{Timer management}
\label{ssec:timer}

Applications such as connection tracking that dynamically create entries require these entries to expire after some time. For example, for connection tracking, an entry is created for the flow after a TCP SYN or an UDP packet is seen. If after a few hours (for TCP) or after a few seconds (for UDP) no packet has been seen on this flow/connection, the connection is assumed to be expired and closed. All entries related to this connection should be deleted.

The strawman approach could rely on a timer, handled by a dedicated component in the system (e.g., hashed or hierarchical timer wheels~\cite{Varghese1987} or a callout-like interface as in DPDK~\cite{DPDK}). Whenever the timer expires, a callback is called, which accesses the hash table and deletes the corresponding entry. This approach has several drawbacks: \emph{(i)} the key to pass as a parameter to the callback must be duplicated in the timer data structure, thus increasing memory usage, \emph{(ii)} the timer must be updated/reset every time a new packet goes through a given connection thus increasing code complexity and computational cost, and \emph{(iii)} on timer expiration, the callback searches the corresponding key in the hash table to delete it, thus generating memory accesses and consuming memory bandwidth.

As our goal is to eliminate all unnecessary memory accesses, we integrate entry expiration in the hash table rather than using an external timer component. We attach to each entry of the hash table an expiration time. We extend the API of the hash table to support setting/updating the expiration time when inserting or looking up a key. When looking up, only non-expired entries are considered. When inserting, expired entries are overwritten as if the slot was free. Thus, expired entries are lazily deleted by the next insertion thus avoiding unnecessary memory accesses and the computational overhead of executing callbacks.

The main issue with the integration of expiration times in the hash table is that memory is very constrained. As cachelines are 64 bytes on Intel Xeon processors, it is not practical to rely on the usual 32 or 64-bit timestamps, so we use 16-bit timestamps. Indeed, for 8-slots per bucket, 32-bits timestamps would consume half of the cacheline, and 64-bits timestamp would consume the entire cacheline. Using 16-bit timestamps comes with several problems \emph{(i)} the maximum expiration time is short, and \emph{(ii)} overflows, which can revive expired entries, cannot be ignored. 

To solve the first issue, we don't manage time is seconds or milliseconds but we quantize it at a larger granularity. For example, for connection tracking, where connections should expire after some minutes of inactivity, we use a basic time unit of 30 seconds (i.e., a entry with an expiration time of $t_0+2$ will expire 60 seconds after $t_0$). 

To solve the second issue, we don't allow expiration times to take any value between $t_0$ and $t_0+65536$ but restrict them to the range $t_0$ to $t_0+1024$. This still allows expiration times of up to 8 hours in the future. This restriction allows us to distinguish valid entries, expired entries and overflowed (i.e., expired) entries : assuming unsigned 16-bit integers, an entry is non-expired if an only if the difference between the expiration time and the current time is lower than the maximum allowed expiration delay (i.e., 1024). Yet, even with this restriction, a timer could still be revived due to overflow after $64512=65536-1024$ time units. To avoid this, the hash table must be scanned every 537 hours ($\approx 64512$ time units) and any expired entry detected during this scan must be marked as deleted so that they are not revived. This remains very infrequent ensuring that the computational cost associated with this operation is low.

\subsection{Memory Layout}
\label{ssec:memory}
The memory is managed by the processor by cachelines. Accessing a never-accessed cacheline is expensive compared to accessing a value to an already-accessed cacheline (cached in the L1 cache). Indeed, the latency to the L1 cache is 4-5 cycles while the latency to the RAM, when data is not cached, is 150-300 cycles on Intel Xeon processors. It is thus crucial to minimize the number of cachelines accessed when performing a lookup. As a consequence, we organize our structures so that all fields accessed when looking up a bucket are in the same cacheline. This influenced design choices such as the use of a relatively small 64-bit bloom filter, or the use of 16-bit timers with limited precision.

Modern processors support SIMD instructions that perform an identical operation on multiple data elements in a single instruction. We leverage this to read the memory, compare hashes and check timers. The SIMD unit (SSE instructions) performs operations on 128-bit registers. As shown on Figure~\ref{plots:ratio_in_secondary}, cuckoo hash table with 8 slots per bucket allow higher load factors, and better performance. We thus use 8 16-bits tags (128 bits), and 8 16-bits timers (128 bits).

The memory layout we use for Cuckoo++ with or without timers is shown on Figure~\ref{fig:bucket_memory}. The data accessed during lookups is shaded. Each bucket stores 8 entries. We store 8 16-bits tags derived from the main hashes, 8 16-bit values corresponding to the expiration time, a bloom filter with a counter, a few 8-bit masks that are used to mark an entry as free or busy, or that are used temporarily when searching for a cuckoo path during insertion. 

When timers are used, the alternative hashes cannot all be stored on the same cacheline. Yet, these hashes are seldomly used. They are needed only for inserts when the primary and the secondary bucket are full, to search a cuckoo path that can free space in buckets. Thus this has no impact on lookup performance.

Figure~\ref{fig:bucket_memory} also shows the memory layout used in DPDK 17.05 hash tables, a widespread hash table implementation supporting large keys and values, and batched lookups. Notable differences between Cuckoo++ and DPDK are: \emph{(i)} the use of only 16-bit tags rather than complete 32-bit main hashes allowing to save room in the cacheline and to use 128-bit SIMD instructions, \emph{(ii)} the use of an implicit index (derived from the bucket index and the position of the bucket) rather than an explicit index for accessing the keys and values which allows to save a lot of room from the cacheline, room which we use for storing timers and the bloom filter.

Cuckoo++ memory layout can be reused for implementing other variants of high-performance Cuckoo hash-table supporting large keys and values on general purpose processors. Our pessimistic cuckoo and optimistic cuckoo implementations used in the study of Section~\ref{sec:study} use the same memory layout as Cuckoo++ with padding instead of bloom filters and moved counters. Horton tables introduced in Section~\ref{sec:eval} do not need the bloom filter nor the moved counter but these are replaced by the remap array which is 63 bits long.

\section{Evaluation}
\label{sec:eval}

\paragraph{Hardware}
Our experiments are carried out on a dual-socket Dell R630 server equipped with two Intel Xeon E5-2640 v4 (10 cores per socket, 2.4 Ghz). The memory installed in the server is DDR4 2133Mhz, with 4 DIMMS per socket for a total of 128 GB (i.e., 8 x 16 GB allowing the processor to operate in quad-channel mode).
The server is configured in \emph{performance} mode, hyper-threading is disabled, and memory snoop mode is set to \emph{home snooping}. For all experiments, memory is allocated on the local NUMA node, and the cores are allocated alternately on each socket.

\paragraph{Common parameters}
We use CityHash 64~\cite{Pike2011} which outputs 64-bit hashes. All keys are randomly generated. To match the behavior of realistic applications, for mixed lookups (e.g., negative lookup rate of 0.2), the number and the position of negative keys in a batch are not constant: the negative lookup rate is the average. We evaluate implementations with 128-bit keys and 128-bit values (32 bytes).

\paragraph{Implementations evaluated}\textcolor{white}{.}\\
\textbf{DPDK (version 17.05)} We used the hash table implementation of DPDK~\cite{DPDK}, the de-facto standard library for high performance packet processing, and the only general-purpose hash table implementation that supports batched lookups. To support our use cases, we increased the amount of data that can be stored in the hash table from 64 bits to 128 bits so as to store data and avoid storing a pointer to data in the hash table with data stored externally. This change has no impact on the implementation's performance as the key and the data (256 bits) still fit a single cache line. This implementation takes a pessimistic approach and always prefetch both the primary and the secondary bucket.\\
\textbf{Cuckoo++} Our implementation of Cuckoo++ as described in Section~\ref{sec:main} with optional timers disabled.\\
\textbf{Cuckoo++ w/ timer} Our implementation of Cuckoo++ as described in Section~\ref{sec:main} with optional timers enabled.\\
\textbf{Optimistic Cuckoo} We cannot compare directly against CuckooSwitch~\cite{Zhou2013} for it is tailored for small keys (48 bits) and very small values (16 bits). We however implement a batching strategy similar to CuckooSwitch. To allow a precise evaluation of the benefits of the algorithmic changes, this implementation shares most of its code with Cuckoo++. This ensures that the performance differences come mostly from the algorithmic changes we introduce and not from differences in implementation. It allows to determine the gains obtained from the sole introduction of the bloom filter.\\
\textbf{Pessimistic Cuckoo} Our implementation of a pessimistic strategy similar to DPDK's one. Similarly to optimistic cuckoo, this implementation shares most of its code with Cuckoo++.\\
\textbf{Horton} Horton tables~\cite{Breslow2016} are a modification of bucketized cuckoo hash tables that also aims at improving the performance for negative lookups. In Horton tables, buckets are augmented with a remap array indexed by a tag. If an entry is not stored in its primary bucket, it might be stored in one of its several secondary buckets as determined by the value stored in the remap array at position tag. If this value is zero, we know that no entry with such tag has been remapped, so we can avoid reading a secondary bucket. Horton hash tables have been designed for small keys (32 bits) and values (31 bits). They were implemented and evaluated only on GPUs, while networking applications run mostly on CPUs. Thus, we cannot compare against standard Horton tables. In order to evaluate the algorithmic changes introduced in Horton tables and compare them to the ones we introduce with Cuckoo++, we implement a variant of Horton tables for CPUs that leverages the optimized code-base and memory layout of Cuckoo++ (see Figure~\ref{fig:bucket_memory}) and that stores the Horton remap array in place of the bloom filter. Since our memory layout is less constrained in term of memory usage, we drop the notion of type A (8 slots) and type B buckets (7 slots) ~\cite{Breslow2016} so as to avoid branching which is detrimental to performance on CPUs. We also implement lookup in batches as required by packet processing applications. To our knowledge, this is the first design and evaluation of Horton tables on CPUs, for larger keys and values, and supporting batching.

\begin{figure*}[t]
	\centering%
	\small
	\begin{tikzpicture}[baseline, trim axis left]
\begin{axis} [
  perf plot insert,
   title={(a) Small table (512k capacity)},
  legend columns=6,
  legend to name=graphs-legend
]

\addplot+
    table[x=load_factor, y=UNCOND_op_rate] {graphs/inserts_x_load-512k-invalid0.gnuplot};
  \addplot+
    table[x=load_factor, y=COND_op_rate] {graphs/inserts_x_load-512k-invalid0.gnuplot};
  \addplot+
    table[x=load_factor, y=DPDK_1702_op_rate] {graphs/inserts_x_load-512k-invalid0.gnuplot};
  \addplot+
    table[x=load_factor, y=HORTON_op_rate] {graphs/inserts_x_load-512k-invalid0.gnuplot};    
  \addplot+
    table[x=load_factor, y=BLOOM_op_rate] {graphs/inserts_x_load-512k-invalid0.gnuplot};
  \addplot+
    table[x=load_factor, y=LAZY_BLOOM_op_rate] {graphs/inserts_x_load-512k-invalid0.gnuplot};

\legend{
    Pessimistic Cuckoo,
    Optimistic Cuckoo,
    DPDK,
    Horton,
    Cuckoo++,
    Cuckoo++ w/ timers
  }

\end{axis}
\end{tikzpicture}%
\begin{tikzpicture}[baseline]
\begin{axis} [
  perf plot insert,
     title={(b) Large table (32M capacity)},
]

\addplot+
    table[x=load_factor, y=UNCOND_op_rate] {graphs/inserts_x_load-invalid0.gnuplot};
  \addplot+
    table[x=load_factor, y=COND_op_rate] {graphs/inserts_x_load-invalid0.gnuplot};
  \addplot+
    table[x=load_factor, y=DPDK_1702_op_rate] {graphs/inserts_x_load-invalid0.gnuplot};
  \addplot+
    table[x=load_factor, y=HORTON_op_rate] {graphs/inserts_x_load-invalid0.gnuplot};
  \addplot+
    table[x=load_factor, y=BLOOM_op_rate] {graphs/inserts_x_load-invalid0.gnuplot};
  \addplot+
    table[x=load_factor, y=LAZY_BLOOM_op_rate] {graphs/inserts_x_load-invalid0.gnuplot};
\end{axis}
\end{tikzpicture}
\begin{tikzpicture}[baseline, trim axis right]
\begin{axis} [
  perf plot insert,
  title={(c) Very large table (128M capacity)},
]

\addplot+
    table[x=load_factor, y=UNCOND_op_rate] {graphs/inserts_x_load-128m-invalid0.gnuplot};
  \addplot+
    table[x=load_factor, y=COND_op_rate] {graphs/inserts_x_load-128m-invalid0.gnuplot};
  \addplot+
	  table[x=load_factor, y=DPDK_1702_op_rate] {graphs/inserts_x_load-128m-invalid0.gnuplot};
  \addplot+
    table[x=load_factor, y=HORTON_op_rate] {graphs/inserts_x_load-128m-invalid0.gnuplot};
  \addplot+
    table[x=load_factor, y=BLOOM_op_rate] {graphs/inserts_x_load-128m-invalid0.gnuplot};
  \addplot+
    table[x=load_factor, y=LAZY_BLOOM_op_rate] {graphs/inserts_x_load-128m-invalid0.gnuplot};
\end{axis}
\end{tikzpicture}\\
\pgfplotslegendfromname{graphs-legend}%
	\normalsize
	\caption{Performance for insertions up to the given load factor for various table sizes.}%
	\label{plots:perf_inserts}%
\end{figure*}
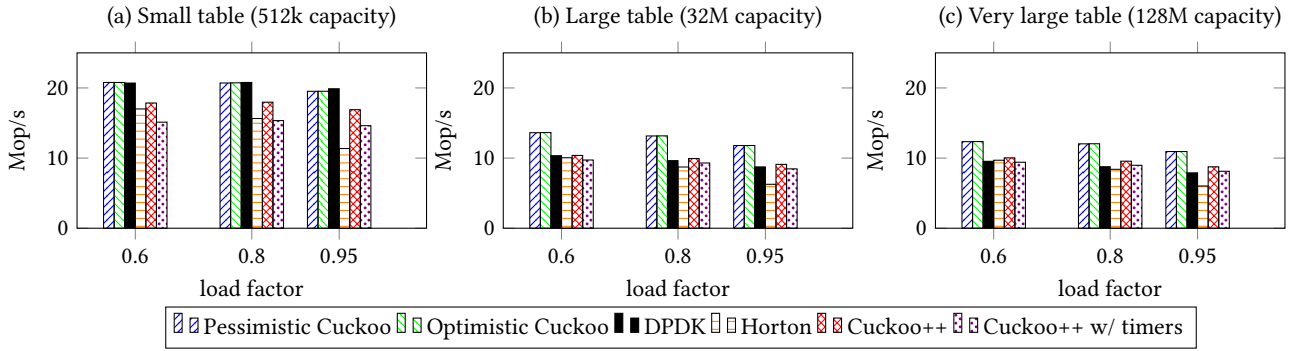

\paragraph{Hash table capacity.} We evaluate the performance for different hash table capacities.\\
\textbf{Small Tables (512K entries)} Our implementation is designed for large tables and thus the minimal capacity supported is 524288 entries (24 MB at 48 bytes per entry). Such table is small enough to remain partially in the L3 cache. \\
\textbf{Large Tables (32M entries)}	Larger tables will not fit the L3 cache anymore. The accesses to the bucket will thus generate L3 cache misses and will have to access the memory with longer latency and lower bandwidth \\
\textbf{Very Large Tables (128M entries )} The last configuration we evaluate considers very large tables. These tables exceed 4GB, the maximum amount of memory that can be addressed by Intel Xeons without generating TLB misses.

\paragraph{Metrics}
The two metrics of interests are: \emph{(i)} insert performance since hash tables used for connection tracking are dynamic and evolve over time, \emph{(ii)} batched lookup performance since high-performance networking applications process batches of packets.

Our primary focus is on improving batched lookup performance. Indeed, we expect lookups to dominate in many workloads, and to become increasingly dominant in the future. Indeed, the number of packets per flow is increasing due to the trend of having longer lived connections through the adoption of HTTP 2.0 and the increase in video traffic~\cite{Qian2012}. In 1997, 20 packets per flow were observed~\cite{Thompson1997}; in 2000-2005, 26 to 32 packets per flow were observed~\cite{Kim2006, Muscariello2006}; and in recent years 2014 to 2016, 81 to 114 packets per flow were observed~\cite{Bocchi2016, Velan2016}.

Beside these performance metrics, we also consider implementation complexity (measured in number of lines of code) and the overhead in memory usage.

\subsection{Implementation complexity}
We evaluate the implementation complexity  by measuring the number of lines of codes (excluding comments and white lines). In Table~\ref{tab:codesize}, we report the number of lines for each variant. The algorithmic changes in Cuckoo++ add 5\% of code to the baseline implementation to implement the bloom filter, its maintenance, and its use to prune accesses to the secondary bucket. When comparing to optimistic cuckoo, which is the implementation having the best performance for positive lookups, the addition code is only 2\%. When compared to Horton, Cuckoo++ require more than twice less additional code, thanks to the simpler algorithm. Timer management is also a minor complexity increase with less than 5\% increase in code size. Overall, all improvements can easily be integrated into existing cuckoo hash tables, except for Horton because it alters the way the secondary bucket is indexed.

\begin{table}[!h]
	\caption{Number of lines of codes per implementation}%
	\label{tab:codesize}%
	\small%
	\begin{tabular}{|c|c|c|}%
	\hline 
	& No timer & Timer \\ 

	\hline
	Pessimistic-Cuckoo & \textbf{850} & 881 (+31)\\
	\hline 
	Optimistic-Cuckoo & 871 (+21) & 902 (+52) \\
	\hline 
	Horton & 954 (+94) & 985 (+135)\\
	\hline 
	Cuckoo++ & 892  (+42) & 923 (+73) \\
	\hline 
\end{tabular} 
	\normalsize
\end{table}

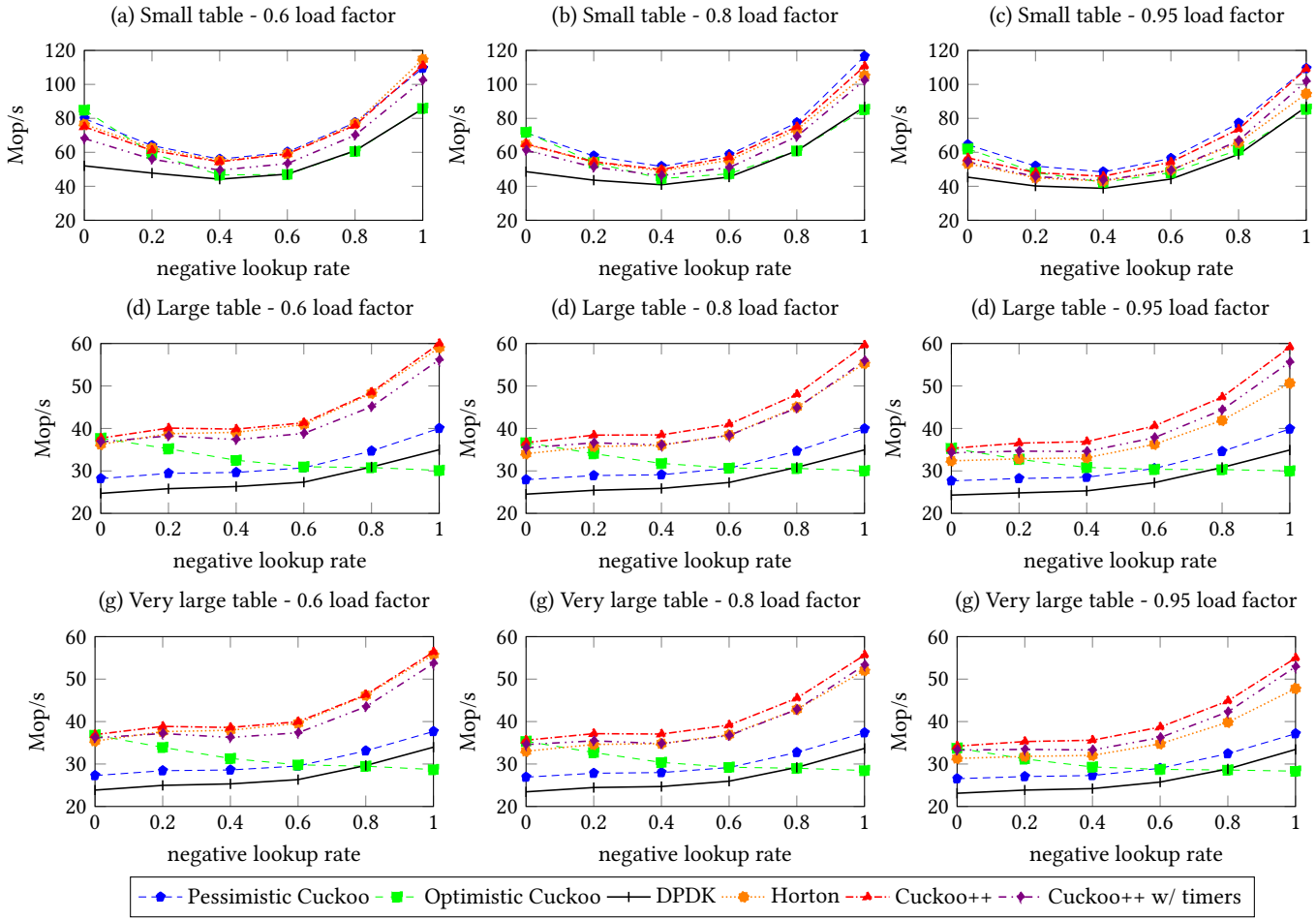
\begin{figure*}
	\centering
	\small
	\begin{tikzpicture}[baseline, trim axis left]
\begin{axis} [
  perf negative plot,
  ymax=120,
  title={\shortstack{(a) Small table  - 0.6 load factor}},
]

  \addplot+
    table[x=invalid_rate, y=UNCOND_op_rate] {graphs/x_invalid-512k-load0.6.gnuplot};
  \addplot+
    table[x=invalid_rate, y=COND_op_rate] {graphs/x_invalid-512k-load0.6.gnuplot};
  \addplot+
 table[x=invalid_rate, y=DPDK_1702_op_rate] {graphs/x_invalid-512k-load0.6.gnuplot};
  \addplot+
    table[x=invalid_rate, y=HORTON_op_rate] {graphs/x_invalid-512k-load0.6.gnuplot};
  \addplot+
    table[x=invalid_rate, y=BLOOM_op_rate] {graphs/x_invalid-512k-load0.6.gnuplot};
  \addplot+
    table[x=invalid_rate, y=LAZY_BLOOM_op_rate] {graphs/x_invalid-512k-load0.6.gnuplot};

\end{axis}
\end{tikzpicture}
\begin{tikzpicture}[baseline]
\begin{axis} [
  perf negative plot,
  ymax=120,
  title={\shortstack{(b) Small table  - 0.8 load factor}}
]
  \addplot+
    table[x=invalid_rate, y=UNCOND_op_rate] {graphs/x_invalid-512k-load0.8.gnuplot};
  \addplot+
    table[x=invalid_rate, y=COND_op_rate] {graphs/x_invalid-512k-load0.8.gnuplot};
      \addplot+
    table[x=invalid_rate, y=DPDK_1702_op_rate] {graphs/x_invalid-512k-load0.8.gnuplot};
  \addplot+
    table[x=invalid_rate, y=HORTON_op_rate] {graphs/x_invalid-512k-load0.8.gnuplot};
  \addplot+
    table[x=invalid_rate, y=BLOOM_op_rate] {graphs/x_invalid-512k-load0.8.gnuplot};
  \addplot+
    table[x=invalid_rate, y=LAZY_BLOOM_op_rate] {graphs/x_invalid-512k-load0.8.gnuplot};
\end{axis}
\end{tikzpicture}
\begin{tikzpicture}[baseline, trim axis right]
\begin{axis} [
  perf negative plot,
  ymax=120,
  title={\shortstack{(c) Small table  - 0.95 load factor}}
]

  \addplot+
    table[x=invalid_rate, y=UNCOND_op_rate] {graphs/x_invalid-512k-load0.95.gnuplot};
  \addplot+
    table[x=invalid_rate, y=COND_op_rate] {graphs/x_invalid-512k-load0.95.gnuplot};
  \addplot+
  table[x=invalid_rate, y=DPDK_1702_op_rate] {graphs/x_invalid-512k-load0.95.gnuplot};
  \addplot+
    table[x=invalid_rate, y=HORTON_op_rate] {graphs/x_invalid-512k-load0.95.gnuplot};
  \addplot+
    table[x=invalid_rate, y=BLOOM_op_rate] {graphs/x_invalid-512k-load0.95.gnuplot};
  \addplot+
    table[x=invalid_rate, y=LAZY_BLOOM_op_rate] {graphs/x_invalid-512k-load0.95.gnuplot};
\end{axis}
\end{tikzpicture}\\
	\begin{tikzpicture}[baseline, trim axis left]
\begin{axis} [
  perf negative plot,
  title={\shortstack{(d) Large table  - 0.6 load factor}},
]

  \addplot+
    table[x=invalid_rate, y=UNCOND_op_rate] {graphs/x_invalid-load0.6.gnuplot};
  \addplot+
    table[x=invalid_rate, y=COND_op_rate] {graphs/x_invalid-load0.6.gnuplot};
  \addplot+
    table[x=invalid_rate, y=DPDK_1702_op_rate] {graphs/x_invalid-load0.6.gnuplot};
  \addplot+
	table[x=invalid_rate, y=HORTON_op_rate] {graphs/x_invalid-load0.6.gnuplot};
  \addplot+
    table[x=invalid_rate, y=BLOOM_op_rate] {graphs/x_invalid-load0.6.gnuplot};
  \addplot+
    table[x=invalid_rate, y=LAZY_BLOOM_op_rate] {graphs/x_invalid-load0.6.gnuplot};
\end{axis}
\end{tikzpicture}%
\begin{tikzpicture}[baseline]
\begin{axis} [
  perf negative plot,
  title={\shortstack{(d) Large table  - 0.8 load factor}},
]
  \addplot+
    table[x=invalid_rate, y=UNCOND_op_rate] {graphs/x_invalid-load0.8.gnuplot};
  \addplot+
    table[x=invalid_rate, y=COND_op_rate] {graphs/x_invalid-load0.8.gnuplot};
  \addplot+
    table[x=invalid_rate, y=DPDK_1702_op_rate] {graphs/x_invalid-load0.8.gnuplot};
  \addplot+
	table[x=invalid_rate, y=HORTON_op_rate] {graphs/x_invalid-load0.8.gnuplot};
  \addplot+
    table[x=invalid_rate, y=BLOOM_op_rate] {graphs/x_invalid-load0.8.gnuplot};
  \addplot+
    table[x=invalid_rate, y=LAZY_BLOOM_op_rate] {graphs/x_invalid-load0.8.gnuplot};
\end{axis}
\end{tikzpicture}%
\begin{tikzpicture}[baseline, trim axis right]
\begin{axis} [
  perf negative plot,
  title={\shortstack{(d) Large table  - 0.95 load factor}},
]

  \addplot+
    table[x=invalid_rate, y=UNCOND_op_rate] {graphs/x_invalid-load0.95.gnuplot};
  \addplot+
    table[x=invalid_rate, y=COND_op_rate] {graphs/x_invalid-load0.95.gnuplot};
  \addplot+
    table[x=invalid_rate, y=DPDK_1702_op_rate] {graphs/x_invalid-load0.95.gnuplot};
  \addplot+
	table[x=invalid_rate, y=HORTON_op_rate] {graphs/x_invalid-load0.95.gnuplot};
  \addplot+
    table[x=invalid_rate, y=BLOOM_op_rate] {graphs/x_invalid-load0.95.gnuplot};
  \addplot+
    table[x=invalid_rate, y=LAZY_BLOOM_op_rate] {graphs/x_invalid-load0.95.gnuplot};
\end{axis}
\end{tikzpicture}\\
	\begin{tikzpicture}[baseline, trim axis left]
\begin{axis} [
  perf negative plot,
  title={\shortstack{(g) Very large table  - 0.6 load factor}},
  legend columns=6,
  legend to name=graphs-legend,
]

  \addplot+
    table[x=invalid_rate, y=UNCOND_op_rate] {graphs/x_invalid-128m-load0.6.gnuplot};
  \addplot+
    table[x=invalid_rate, y=COND_op_rate] {graphs/x_invalid-128m-load0.6.gnuplot};
  \addplot+
	table[x=invalid_rate, y=DPDK_1702_op_rate] {graphs/x_invalid-128m-load0.6.gnuplot};
  \addplot+
    table[x=invalid_rate, y=HORTON_op_rate] {graphs/x_invalid-128m-load0.6.gnuplot};
  \addplot+
    table[x=invalid_rate, y=BLOOM_op_rate] {graphs/x_invalid-128m-load0.6.gnuplot};
  \addplot+
    table[x=invalid_rate, y=LAZY_BLOOM_op_rate] {graphs/x_invalid-128m-load0.6.gnuplot};

  \legend{
    Pessimistic Cuckoo,
    Optimistic Cuckoo,
    DPDK,
    Horton,
    Cuckoo++,
    Cuckoo++ w/ timers
  }
\end{axis}
\end{tikzpicture}
\begin{tikzpicture}[baseline]
\begin{axis} [
  perf negative plot,
  title={\shortstack{(g) Very large table  - 0.8 load factor}},
]
  \addplot+
    table[x=invalid_rate, y=UNCOND_op_rate] {graphs/x_invalid-128m-load0.8.gnuplot};
  \addplot+
    table[x=invalid_rate, y=COND_op_rate] {graphs/x_invalid-128m-load0.8.gnuplot};
  \addplot+
    table[x=invalid_rate, y=DPDK_1702_op_rate] {graphs/x_invalid-128m-load0.8.gnuplot};
  \addplot+
    table[x=invalid_rate, y=HORTON_op_rate] {graphs/x_invalid-128m-load0.8.gnuplot};
  \addplot+
    table[x=invalid_rate, y=BLOOM_op_rate] {graphs/x_invalid-128m-load0.8.gnuplot};
  \addplot+
    table[x=invalid_rate, y=LAZY_BLOOM_op_rate] {graphs/x_invalid-128m-load0.8.gnuplot};
\end{axis}
\end{tikzpicture}
\begin{tikzpicture}[baseline, trim axis right]
\begin{axis} [
  perf negative plot,
  title={\shortstack{(g) Very large table  - 0.95 load factor}},
]

  \addplot+
    table[x=invalid_rate, y=UNCOND_op_rate] {graphs/x_invalid-128m-load0.95.gnuplot};
  \addplot+
    table[x=invalid_rate, y=COND_op_rate] {graphs/x_invalid-128m-load0.95.gnuplot};
  \addplot+
	 table[x=invalid_rate, y=DPDK_1702_op_rate] {graphs/x_invalid-128m-load0.95.gnuplot};
  \addplot+
    table[x=invalid_rate, y=HORTON_op_rate] {graphs/x_invalid-128m-load0.95.gnuplot};
  \addplot+
    table[x=invalid_rate, y=BLOOM_op_rate] {graphs/x_invalid-128m-load0.95.gnuplot};
  \addplot+
    table[x=invalid_rate, y=LAZY_BLOOM_op_rate] {graphs/x_invalid-128m-load0.95.gnuplot};
\end{axis}
\end{tikzpicture}\\
\pgfplotslegendfromname{graphs-legend}
	\normalsize
	\caption{Performance for lookups on a single core with different capacities (rows) and different load factors (columns).}
	\label{plots:perf_lookups}
\end{figure*}

\subsection{Evaluation of insert performance}

Figure~\ref{plots:perf_inserts} plots the number of insertions performed per second in a hash table for a given load factor. We show results for the different hash table capacities.

For small tables, our implementations of optimistic cuckoo and pessimistic cuckoo achieve the highest performance together with DPDK. Cuckoo++ has a slight overhead due to the management of the bloom filter. Cuckoo++ with timer pays an additional overhead for timer management. Yet, this overhead remains negligible and is easily offset as it avoids to manage timers externally. 

For large and very large tables, the performance of Cuckoo++ and DPDK is similar and just slightly lower than pessimistic cuckoo and optimistic cuckoo. It decreases only slightly as the load factor increases. 

For all table sizes, Horton table performance decreases significantly as the load factor increases. Indeed, the insertion procedure in Horton tables is much more complex, especially as the primary bucket is full, since it requires additional hash computations and cacheline accesses to search the least loaded among several potential secondary buckets.

\subsection{Evaluation of lookup performance}

We evaluate the performance of batched lookups, for different hash table capacity, for different load factors, and for different ratio of negative lookups. The results are reported in Figure~\ref{plots:perf_lookups}.

\subsubsection{Design choices}
Our design goal was to offer the best of the two alternatives: having good performance for positive and negative lookups. We thus compare the performance of Cuckoo++ to optimistic cuckoo and pessimistic cuckoo. Cuckoo++ outperforms both, thus achieving our design goal. It performs similarly to optimistic cuckoo for low negative lookup rate. Indeed, by construction, both generate the same number of memory accesses, and the cost of using the bloom filter is negligible. It outperforms both optimistic and pessimistic cuckoo for high negative lookup rate: for negative lookups, Cuckoo++ generate a single memory access to the primary bucket, while standard cuckoo generate two memory accesses to the primary and secondary buckets, with more or less efficient prefetches.

The behavior of Cuckoo++ tends not to be impacted by the load factor. However, Cuckoo++ behaves quite differently depending on the size of the tables. For small tables, performance is higher (as for DPDK, optimistic cuckoo and pessimistic cuckoo). This is because memory accesses have a relatively low latency since the table can persist in the L3 cache. We also observe that for mixed workload (0.5 negative lookup rate), the performance is lower. This is because in this case, the branch predictor of the processor is unable to predict correctly the code to execute. For larger tables (32M and 128M capacity), the performance is similar, interestingly there is little differences between 32M capacity tables (which do not generate TLB misses) and 128M capacity tables (which do generate TLB misses). Also, for both sides, the branch prediction failures are amortized by the memory latency and thus we don't have U-shaped curves.

\subsubsection{Comparison with alternatives}
First, the baseline performance of our implementation is improved over DPDK, even when using the same prefetch strategy as DPDK (i.e., pessimistic cuckoo). This is enabled by our optimized memory layout that is more compact and that allows leveraging SSE to match hashes. Second, Cuckoo++ outperforms DPDK for all configurations, similarly to the way cuckoo++ outperforms pessimistic cuckoo.  Indeed, by avoiding the un-necessary prefetch of the secondary bucket for positive lookups, the performance is improved since this prefetch is useless more than 80\% of the time. Similarly, for negative lookups, the performance is also improved significantly as only the primary bucket needs to be accessed.

We evaluate the effectiveness of Horton tables on CPUs (the original paper targeted GPUs and restricted keys/values sizes). Horton tables perform reasonably for low load factors providing performance just below optimistic cuckoo for positive lookups. Yet, as the load factor increases, the performance gap increases and Horton tables are outperformed by optimistic cuckoo. Thus, horton tables cannot be used as a universal replacement for all settings. Cuckoo++ outperforms horton tables for all settings (all tables sizes and all load factors). Also note that contrary to horton tables, cuckoo++ does not alter the theoretical guarantees of cuckoo tables, and cuckoo++ is simpler to implement than Horton (see Table~\ref{tab:codesize}).

To explain the improved performance with Cuckoo++, we compute the expected false positive rate (i.e., when the bloom filter falsely indicates that the secondary bucket needs to be checked) for various load factors for a hash table of capacity 32M. Horton tables also provide a hint that allows to avoid accessing the secondary bucket (i.e., whenever the remap array entry is set to zero). Thus, we also compute the false positive rate associated with the remap array of Horton tables for the same setting. The false positive rates are reported in Table~\ref{tab:fpr}.
 
 \begin{table}[t]
 	\caption{False Positive Rate (FPR) with Bloom Filter (Cuckoo++) and Remap Array (Horton tables)}
 	\label{tab:fpr}
 	\small
 	\begin{tabular}{|c|c|c|}
 		\hline 
 		Load factor & FPR (Cuckoo++) & FPR (Horton)  \\ 
 		\hline 
 		$l = 0.6$ &  0.0002 & 0.004\\ 
 		\hline 
 		$l = 0.8$ &  0.001 & 0.02 \\ 
 		\hline 
 		$l = 0.95$ & 0.003 & 0.05 \\ 
 		\hline
 	\end{tabular} 
 \normalsize
 \end{table}
 
For all load factors, the false positive rate for Cuckoo++ is low (less than 0.3\%) ensuring good pruning performance and avoiding most unnecessary accesses to the secondary bucket.
Interestingly, due to the differences in structures the false positive rate for Cuckoo++ is much lower than for Horton tables. This, combined with the more complex code, explains the difference in performance between Cuckoo++ and Horton tables especially as the load factor increases.

As a conclusion, Cuckoo++ is a good alternative to both optimistic and pessimistic cuckoo  hash table implementations (including DPDK). Contrary to Horton tables, it consistently improves performance even for high load factors. Cuckoo++ is the only implementation that offers excellent batched lookup performance in all cases without significantly decreasing the insert performance. Moreover, Cuckoo++ has a moderate implementation complexity and very moderate memory overhead.

\subsubsection{Overhead of timer management}
Figure~\ref{plots:perf_lookups} also plots the performance of Cuckoo++ with timers alongside the performance of Cuckoo++ (without timers). For all settings, the impact on performance remains low (< 5\%). This is much lower than relying on external timers and performing explicit deletions in the hash table. Even with the timer functionality enabled, Cuckoo++ remains among the best implementations in term of performance.

\subsection{Performance in a Multi Core Setting}
While our hash table is designed for use in a share nothing architecture (i.e., with each core running its own thread with its private hash table), it is important to evaluate its performance in a multi-core setting. Indeed, hash tables are memory-intensive and put a lot of pressure on the shared hardware resources of the CPU such as the L3 cache, or the memory controller.

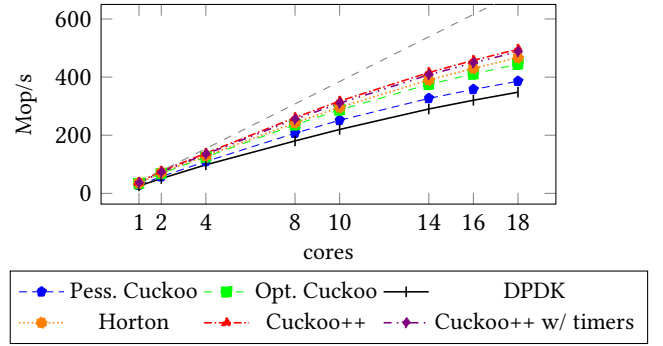
\begin{figure}[t]
	\centering
	\begin{tikzpicture}[baseline, trim axis left, trim axis right]
\begin{axis} [
  cycle list name = {customcyclelist},
  single plot,
  xtick = {1,2,4,8,10,14,16,18},
  xlabel = cores,
  ylabel = Mop/s,
  ymax = 650,
  legend to name=multicore-legend,
  legend columns=3,
]
  \addplot+
    table[x=core_count, y=UNCOND_global_op_rate] {graphs/x_cores.gnuplot};
  \addplot+
    table[x=core_count, y=COND_global_op_rate] {graphs/x_cores.gnuplot};
  \addplot+
    table[x=core_count, y=DPDK_1702_global_op_rate] {graphs/x_cores.gnuplot};
  \addplot+
	table[x=core_count, y=HORTON_global_op_rate] {graphs/x_cores.gnuplot};
  \addplot+
    table[x=core_count, y=BLOOM_global_op_rate] {graphs/x_cores.gnuplot};
  \addplot+
    table[x=core_count, y=LAZY_BLOOM_global_op_rate] {graphs/x_cores.gnuplot};
  \addplot+ [gray, dashed, mark=none, samples at={1,18}] {38.47 * x};
  
    \legend{
    Pess. Cuckoo,
    Opt. Cuckoo,
    DPDK,
    Horton,
    Cuckoo++,
    Cuckoo++ w/ timers
  }
\end{axis}
\end{tikzpicture}\\
\pgfplotslegendfromname{multicore-legend}
	\caption{Performance for lookups on multiple cores. Capacity of 32M entries per core, load factor of 0.8 and negative lookup rate of 0.2. The grey dashed line shows ideal (i.e., linear) scaling relative to Cuckoo++.}
	\label{plots:perf_multicore}
\end{figure}

We evaluate their performance from 1 to 18 threads running on the 20 cores of our dual socket Xeon E5-2640v4 (we leave one core on each socket for the OS). We show the results on Figure~\ref{plots:perf_multicore}. The scaling is good, which is coherent with our design to adopt a share-nothing architecture. Yet, even if the data structures are independent and the cores do not synchronize, we observe that the scaling is not linear due to the pressure on shared units of the CPU such as the L3 cache and the memory controller. Cuckoo++ achieves 490M operations per second on 18 cores. The total capacity of the hash tables is 576M (18 x 32M).

\subsection{Memory usage}

Memory usage derives directly from the memory layout we use. Since all our implementations (Horton, Optimistic Cuckoo, Pessimistic Cuckoo and Cuckoo++) share the same memory layout, they all have the same memory requirements. 
We compare it to Cuckoo++ with timer and DPDK. Notice that buckets are aligned to cachelines which means that padding is added to the structures depicted in Figure~\ref{fig:bucket_memory}. Despite the additional metadata needed for Cuckoo++ (i.e., the bloom filter), Cuckoo++ has a lower memory overhead than the implementation of DPDK.

\begin{center}
	\small
	\begin{tabular}{|c|c|c|}
		\hline 
		Variant & Bytes/Entry & Overhead\\ 	
		\hline 
		Cuckoo++ & 48 & 50\% \\ 
		\hline 
		Cuckoo++ with timer & 64 & 100\% \\ 
		\hline 
		DPDK & 64  & 100\% \\ 
		\hline 
	\end{tabular} 
	\normalsize
\end{center}

\section{Related Work}
Optimizing the performance for negative lookups is a recurring concern for open addressing hash tables~\cite{Celis1985,Celis1989,Herlihy2008}. Cuckoo hash tables~\cite{Pagh2004,Panigrahy2005}, especially BCHT~\cite{Dietzfelbinger2007} represent a breakthrough by moving most complexity related to collisions to the insert procedure allowing for efficient lookups: they lower the total number of memory accesses, in both average and worst case. Our goal, similarly to Horton tables~\cite{Breslow2016}, is to go beyond these, exploiting the fact that modern CPUs manage the memory by 64B chunks allowing buckets to store metadata (i.e., the bloom filter) in addition to the slots. Doing so, we have been able to reduce the number of memory accesses from 2 to 1, resulting in improved performance. Horton tables~\cite{Breslow2016} have a similar approach, but they are less efficient at pruning accesses (see FPR reported in Table~\ref{tab:fpr}), and the computational cost of the changes offsets the benefits for high load factors.

In our design, we use bloom filters~\cite{Bloom1970} instead of counting bloom filters~\cite{Fan2000} or cuckoo filters~\cite{Fan2014} that support deletion or have improved FPR. We needed a
a probabilistic filter with very simple lookup procedure : lookups in bloom filters can be implemented with an equal and a bitwise and. Also, a bloom filter extended with a global counter is a more compact solution than counting bloom filters supporting deletion. As shown on Figure~\ref{fig:distrib_moved}, the occupancy remains low, allowing the bloom filter to be reset often enough even if it does not support deletion: we thus favored computational cost and compacity in our design. 

MemC3~\cite{Fan2013} integrates additional functions into the hash table data structure. In the context of memcached, they replace two inter-dependant data structures, the first managing key/value association, and the second keeping track of entries for providing LRU replacement by a single hash table integrating a CLOCK-based LRU eviction. The goal of this integration is to simplify software and to improve performance by reducing overhead, similarly to what we achieve by integrating timers directly into the hash table. Note that, entries are over-written thus providing a form of lazy deletion. Yet, this lazy deletion differs from lazy deletion in open-addressing hash table~\cite{Celis1989} where entries are marked with a tombstone, which still required a memory access. In open addressing hash tables predating cuckoo hash table, these tombstones were useful to avoid costly restructuration of hash tables upon deletion.

Recent works~\cite{Li2014,Zhou2013} consider concurrent read/write accesses to cuckoo hash tables. When implemented, these require additional hardware support or instructions to check coherency. Our application, which leverages the multi-queue capabilities of modern NICs to run in a share-nothing setting can drop this functionality for improved performance. Yet, our modifications to cuckoo hash table are orthogonal to these works, and thus both could be integrated together for improved performance with concurrent accesses.

An alternative solution to store association between connections and state is perfect hashing, as used in ScaleBricks~\cite{Zhou2015}. Perfect hashing allows a compact memory representation and very efficient lookups. The downsides are that it does not answers negatively to lookups but gives a random value, and that insert are more costly. In the context of high-performance packet processing, ScaleBricks~\cite{Zhou2015} achieves 520M lookups per second on 16 cores. Thus, Cuckoo++, at 460M lookups per second on 16 cores, provides an interesting, more flexible, alternative to ScaleBricks whenever the insertion/update rate is higher, or when perfect hashing does not satisfy the requirements of the application (e.g., firewall/NAT must identify non-existing connection, and require to frequently create new associations).

\section{Conclusion}

We implemented an highly efficient cuckoo hash table supporting several strategies (i.e., optmisitic or pessimistic) and evaluate it to show that none is ideal for all workloads. We thus propose Cuckoo++ which adds a bloom filter in the primary bucket, allowing to prune unnecessary accesses to the secondary bucket without requiring expensive computation. Cuckoo++ hash tables have a uniformly good performance when compared to both pessimistic and optimistic implementation, and an improved performance over DPDK and Horton tables for all cases.

We also describe a variant of Cuckoo++ that integrate support for entry expiration directly in the hash table, avoiding the need for external management of timers and the associated overheads. This relies on a new memory layout more compact than DPDK's original one, and on the use of 16-bit timestamp. 

Overall, Cuckoo++ hash tables are a good alternative to existing implementation such as DPDK, and can be used for a wide range of settings (i.e., all capacities, load factor and negative lookup rate). Their support of batched lookups, entry expiration, and good performance for negative lookup rate, makes them well suited for packet processing applications requiring connection tracking.

\begin{acks}
The author is very grateful to Fabien Andr\'e for insightful discussions, comments on the manuscript and support for evaluation. 
\end{acks}

{
	\bibliographystyle{acm}
	\bibliography{cuckoo.bib}
}

 \end{document}